**Infusing equity, diversity, and inclusion throughout our physics curriculum: (Re)defining what it means to be a physicist**

Martha-Elizabeth Baylor[1], Jessica R Hoehn[2], Noah Finkelstein[2]
[1] Department of Physics and Astronomy, Carleton College, Northfield, MN 55057
[2] Department of Physics, University of Colorado, Boulder, Colorado 80309

**Introduction**

Increasingly, the physics community is attending to issues of equity, diversity, and inclusion (EDI), both in language and action[1]–[3]. We are more publicly recognizing our individual responsibilities as physicists to address social injustice and systemic oppression. To this end, our physics classrooms are key levers for action. While we have made great strides in teaching traditional content and habits of mind useful in physics, if we expand our educational practices to more completely reflect the full range of what it means to be a physicist, we can make our classrooms more inclusive and equitable and support the needed diversification of our field. Ultimately, how, and by whom, physics is enacted are part of the discipline itself and ought to be included in our teaching about the professional practices of physicists. We present two complementary approaches for broadening the focus of our classes, share models for enacting these in middle-division physics courses at two distinct institutional types, and provide evidence of positive impacts on students' sense of identity and notions of what it means to be a physicist.

The physics education community has been developing models and practices that make our classrooms more inclusive and explicitly attend to EDI issues and belonging in physics[1], [2], [4]–[8]. Such moves are essential to support students and teachers, and for the discipline itself to bring about much needed systemic change. Borrowing from, and building on, the work of many others[4]–[8], this paper focuses on how EDI can be part of the physics habits of mind that we teach students in middle division physics courses by engaging in extended, explicit discussions and reflections about EDI. Such practices can foster a sense of identity and belonging, which along with appropriate academic support, promote student success[9], [10].

We present two different approaches from two different environments—the Atomic and Nuclear Physics class at Carleton College and the Modern Physics for Engineers class at the University of Colorado (CU) Boulder[*]. These two examples showcase the breadth of opportunities as well as some common features for enacting these essential practices. At Carleton, weekly homework assignments include reflection essays on a variety of topics important to physicists. At CU, students engage in activities and discussions both in and out of class during a two-day unit on representation in science, technology, engineering, and math (STEM). Common among these approaches are themes of focusing on people, building time for reflection, developing and understanding identities in physics, motivating the EDI discussions using the broader physics community as the authority on what physicists care about, and connecting these themes to traditional content of physics as a core, graded component of the course. In this paper, we describe each of our approaches, present some sample outcomes to illustrate the potential impacts for students, and discuss the factors one might consider when determining which approach suits their own local context. The course materials discussed here are available online[11] and we provide additional details about institutional and course contexts as well additional outcomes in the online appendices.

---

[*] For more information on institutional context, see Appendix A.

**Carleton College approach**

Carleton's EDI assignments take place in a 30-person, Sophomore-level course where we have observed that students' experience in this course significantly impacts their decision to major in physics, which they are not allowed to declare until the spring trimester of their second year. The entire course is framed by the theme "Practicing Professionalism" that asks students to consider how every assignment and activity in the course is designed to provide them the opportunity to practice the activities of professional physicists. At the core, the course is designed to address the questions, "What does it mean to do physics?" and "What does it mean to be a physicist?" Though not required for our intervention, this broader professionalism framing naturally allows space in the course to move beyond subject-matter content and skill development (e.g., experimental, computational) to thinking about the importance of EDI, and other non-technical topics. The specific goals of the reflection essay assignment are for students to: (a) examine their own thoughts on what it means to be a physicist and do physics and (b) to read about and respond to EDI-related issues of interest to professional physicists. These non-technical topics are motivated by a 10-minute activity on the first day of class. Each student is given a copy of *APS News*, a compilation of articles that communicate information of value to the physics community. The students are given 5 minutes to look through the newsletter and identify an article whose content does not focus on technical physics topics (e.g., theories, apparatus). The students then briefly introduce themselves and provide the title or topic of the article they identified. As a class, students compile a list of topics physicists care about that are not related to the technical aspects of doing physics, such as: history, awards, leadership, international community, diversity, inclusion, political advocacy, access to physics education, and more. These topics provide a rich array of themes for the reflection essay assignment that are motivated by the larger community of physicists rather than a particular faculty member. This approach to motivating the essays also increases student buy-in and reduces student pushback

In their weekly homework, students engage in a reflection essay exercise. The source-material for the essays are selected readings (e.g., articles, websites), American Institute of Physics (AIP) data, activities inspired by the STEP UP program[7], and/or work by Daane, *et al*[8]. The students must successfully pass nine reflection essays of 500-750 words each that are graded pass/fail based on whether the essays are within the word limit and explicitly engage with the pre-essay content through guiding questions. The instructor reads each of the weekly essays and provides (brief) written feedback on a subset of the essays, with the goal of engaging with each student's work at least twice over the term. The completion of the reflection essay project counts as 5% of the entire course grade. To manage student workload issues, one problem was removed from each homework assignment and replaced with a reflection essay.

Students complete 8 reflection essays over our 9.5-week terms. A selection of these topics include: defining physics and physicists, discussing diversity in physics, exploring collaboration in physics, investigating the Nobel prize, and delving into access and cultural expectations around physics. We provide an example prompt in Figure 1, and each of the specific prompts are provided in the online course materials[11]. By writing these weekly essays, the students begin to build a more realistic physics identity that they can see themselves in. In their initial writings, many students believe real physicists must have a PhD. By the end of the course, many students gain a broader perspective on who does physics (i.e., anyone) and that physicists do more than just solve physics problems. For example, a female student writes, "...I have never been confident that I belong to the physics community. [The reflection essays] taught me to keep challenging myself about the stereotypical views of physicists: it does not have to be white/male/genius/people from academic families/smartest kid in class/etc., all it takes is really a passion for physics." A male student writes, "...My impression of what a physicist does has also changed substantially...I've learned how much more there is to physics than just simply developing models and following the scientific method...and that truly keeping an open mind involves having an awareness for a lot more than just data. "

**Figure 1.** Condensed version of the week 7 reflection prompt (full prompt available online[11]).

> During week 1, you read a statement from Sam Aronson about the importance of diversity in physics. As a result for many decades, the broader physics community has been actively working on trying to improve the representation of individuals from diverse backgrounds and perspectives in the physics community. This week, I want you to think about who does physics and why...consider the following questions:
>
> - When you think about what type of people do physics (e.g., characteristics of a stereotypical physicist, a mentor you look up to, the public face of physics, etc.), what comes to mind? Does that match up with your experience/observations?
> - What are several reasons that you have heard as to why particular groups of people study (and continue to study) physics and others don't? In particular, why do so few women and minorities pursue physics? How does access to high school physics courses impact who pursues physics?
> - Do any of these reasons resonate with your experience? Are any of these reasons inconsistent with your experience?
> - Do you consider yourself the "type of person who does physics" (or whatever your potential major is if you aren't planning to major in physics)? Explain.

Though many students initially resist the reflection essay project, most students ultimately find the assignment valuable. From an anonymous course evaluation, a student comments, "I gained a lot of insights from the reflection essays. I really did hate them at the start of the term but I did not realize how much I have grown mentally after writing all these essays." Sometimes the benefits of the reflection essays come much later, such as in an email from an international student in June 2020. In part she writes, "...today, scientists are practicing things differently to take actions toward eradicating racism (#ShutDownAcademia #ShutDownSTEM). There are resources for us to take concrete actions.  One of them is having discussion about racism and inequality, and I really appreciate that you put assignments about systematic inequality into [the] course even though there are many materials to be covered." This quote shows the long-term impact of placing this type of reflection assignment into a physics course. Appendix B includes exemplar student responses reflecting the impact of this assignment on students' understanding of physicists and the physics community. Moreover, the addition of this reflection essay has not decreased the percentage of students who chose to major in physics after taking this course.

**CU approach**
The course at CU is a Sophomore-level Modern Physics course which enrolls predominantly engineering majors (75-120 students per semester)[†]. Drawing on the work of Daane *et al.*[8], we have incorporated a two-part in-class unit on "What is physics?" and "Who does physics?" (one 75 minute class period for each plus associated readings, homework, and exam questions). Our goals for this unit are for students to: (a) recognize and discuss the role of people in science, (b) identify the lack of diversity in STEM fields and discuss why representation matters, and (c) begin to consider questions of who does science as part of science itself. We believe a supportive and inclusive classroom culture is a necessary condition for successfully addressing these goals; for details about how we strive to foster this environment and the specifics of the implementation, see Ref.[12]).

For part one, students read and discuss a chapter on the nature of science[13] and we spend one class period discussing what it means to create, test, and interpret scientific models and theories[14]. This culminates in a class-wide discussion on the role of people in the process of science (e.g., who creates said models?). For part two, students google "famous physicists" and discuss patterns that they see in the resulting images. The students then practice interpreting graphs and identify the lack of representation in STEM fields using data from APS on participation in physics by race, gender, and LGBTQ identification. After engaging with the data, we ask students why representation matters (if at all). Following our in-class discussions, the students read "Open Letter to SCOTUS from Professional Physicists"[15], and personally reflect on the value of diversity in a physics classroom.

In part one of the unit, students begin to consider the role of people in science, many for the first time. In an online discussion about a reading assignment[13] one student commented, "as students, we were always given equations and diagrams as laws. This [reading] made me think of science as something more approachable than I have always presumed." Throughout the semester, students continue to identify and emphasize the roles of people, collaboration, and community consensus, even when not prompted to do so. For example, in discussing a Scientific American article about the first 100 years of Quantum Mechanics[16], one student remarked, "Science to me is not a competition with right and wrong answers, but is a collaborative effort in which viable explanations are sought after to add to the overall knowledge of the world. The hydrogen disaster is a fine example of scientists building on top of other scientists' work." We also see the students connecting these ideas to their own learning in the course. After the reading and class discussions around the role of people in science, one student wrote about the collaborative nature of science, saying, "it made me realize that working together really is better, and that I shouldn't be afraid to share my ideas, because even if they're wrong, they may help lead me and everyone else to the correct conclusion." In bringing attention to the human elements of science, we create a course where students are empowered to reflect on and construct their own opportunities for learning.

---

[†] This course is similar to the physics majors version of Modern Physics, with slightly more emphasis on conceptual understanding and real-world applications of quantum mechanics. Nothing in this approach would preclude it from being run in the sister course for physics majors.

In part two, we see students engage deeply with the question "why does representation in physics/STEM matter?" During class, students write their own answers to this question on notecards that get distributed anonymously to peers in order to facilitate a class-wide discussion. Common themes in students' responses (across two semesters of this activity) include: leads to diverse ideas and perspectives, diversity leads to better science, sense of belonging, opportunities for individuals, equity, distribution of power/influence, recognition, dismantle prejudice, and access to education/scientific knowledge. Appendix B includes exemplar student responses around each of these themes. Through this activity, students challenge and respond to one another in a productive dialogue, and engage in reflection to identify a variety of important topics.

We first assess the impact of this unit by attending to students' social and emotional reactions. Future work may include additional measures to evaluate whether students are meeting the learning goals of the unit. Students' general reactions to the data interpretation activity and in-class discussion about representation have been largely positive; over two semesters, 75-80% of responses to a weekly feedback survey were positive, 15-20% neutral, and 0-5% negative. Much like at Carleton, our students often express surprise and appreciation for covering these topics in class. The few negative reactions that we have received have primarily included the idea that these topics are important, but they just do not belong in a physics class. We see these responses as being potentially productive and raising awareness for students that these discussions are part of professional physics practice. Further, this unit may be particularly impactful for students from groups that have been historically marginalized or excluded from STEM. In an optional final paper, one student, who self-identified as a woman and member of a racial/ethnic minority group said of the class session on representation, "This class opened my eyes. It may have been one of the most influential classes I have attended to this day." She chose to write her final paper on the topic of diversity in physics and ended with the following: "I took this course to defy odds. I took it to be among the small group of minorities pursuing a degree in physics. I took it to be one of the few women who can proudly call themselves a physicist." In our classes, we have the opportunity to help students develop and sustain their identities as physicists; as illustrated in the preceding quotes, the results can be profound.

**Discussion and conclusions**

We demonstrate that our classrooms can include EDI topics as an *embedded* part of the curriculum, so that all students have the opportunity to engage in a substantive way with EDI as part of the professional practices of physicists. Through these embedded, explicit discussions of EDI, we can positively impact the lives of our students, empower educators, and transform our field for the better. Whether having asynchronous, out of class opportunities (Carleton) or a dedicated in-class unit (CU), we provide students the opportunity to engage with issues of who does physics and what it means to do physics, focusing on people, reflection, and identity, as connected to traditional content of physics. In each approach we emphasize these are discussions occurring in the professional physics community, drawing on resources and data from APS, AIP, AAPT and others. Physics instructors looking to implement these or other curricular materials should do so based on local contexts, opportunities, and constraints. Having discussions with colleagues (and students) can inform which approach makes the most sense in your environment. The Carleton model may be a particularly attractive approach for those who would prefer not to lead these types of sensitive discussions in the classroom, since it allows for offline responses to student reflections, opportunities for engaging with individual students outside of the classroom, and/or in-class discussion if desired. The CU model may be attractive in environments seeking to foster classroom-based discussions and proactively engaging students to respond to each others' perspectives. Both approaches support the inclusion of EDI as part of physics.

For those considering broadening your classroom experiences and opportunities, we applaud you and offer the following recommendations. Start by building an inclusive classroom environment[3], [5]. Then try one of the approaches shared above or in the references below (e.g., reflection essays, readings, classroom discussions). While many of us are not trained in these approaches (or teaching more generally), we can seek opportunities to grow in our ability to facilitate this work. Notably, these class activities come only at a modest cost of traditional content coverage (CU) or time dedicated to practicing traditional content (Carleton), yet they have the potential to greatly expand opportunities for our students and our field. While the overall responses have been very positive, at both Carleton and CU we have used resistance and negative feedback from students to iterate and improve the implementation of our materials. The specifics of implementation and modification will depend on your local context as well as on your positionality and strengths as an instructor.

As we work towards expanding our physics courses in these ways, the value of preparing the next generation of physics students to understand that attending to EDI is part of what it means to be a physicist is the tangible shift we will see in the physics community. Let us choose to embrace this challenging, yet important work of intentionally (re)defining who a physicist is and what a physicist does.

**Acknowledgements**
Thanks to the Carleton and CU Boulder students who have engaged with these classes, contributed to the productive discussions and positive learning environments, and provided valuable feedback that has helped us improve our materials. Additional thanks to Melissa Eblen-Zayas, Director of Carleton's Learning and Teaching Center, and George Cusack, Director of Writing Across the Curriculum for helpful conversations in refining the prompts and the implementation of the reflection activity used at Carleton. A Carleton College Curricular Innovation Grant was used to fund the development of the reflection essay assignment and a Dean's Incubator Grant and a Writing Across the Curriculum Grant funded the development of the SERC Codebook for the Carleton reflection essays. At CU, Julian Gifford, Alexandra Lau, and Alanna Pawlak have contributed to the design and implementation of this course. The work at CU was supported by NSF grant No. 1625824 and a Graduate Research Fellowship. Any opinions, findings, and conclusions or recommendations expressed in this material are those of the authors and do not necessarily reflect the views of the National Science Foundation.

# Appendix A

This appendix includes additional institutional and course context.

**About Carleton College**
Carleton College is a private liberal arts college of 2000 students located in Northfield, MN and operates on the trimester system. The Physics and Astronomy Department graduates an average of 20 physics majors per year. Roughly 30-40% of majors are female-identifying students, 60-70% are male-identifying, 20% are domestic minorities, 70% are white, and 10% are international students. The department has 8 full-time faculty with research programs that cover theoretical/computational physics, experimental physics, and astrophysics. Students are required to take one term of introductory physics, covering Newtonian Mechanics and Special Relativity, before they proceed into the majors sequence of courses at the 200-level. This term of introductory physics does not separate majors from non-majors. Thus, the 200-level courses are the first time our curriculum focuses specifically on the development of our majors. Moreover, students cannot declare their major until their sixth term, so the department does not know until spring term sophomore year which students will declare physics as a major. It is within this context that a reflection essay assignment is introduced into the Fall term 200-level course, Atomic and Nuclear Physics, to build physics identity and introduce issues of equity, diversity, and inclusion in physics.

Atomic and Nuclear Physics is a sophomore-level modern physics course that includes quantum mechanics and its applications to atomic physics and nuclear physics. The course has a required lab component and typically enrolls 30-36 students. Historically, we have seen that students' experience in this class impacts their decision to major in the department. Thus, the faculty are very concerned about the quality of teaching and the student experience in this course.

Course materials related to the reflection essay assignment are available at [link removed for confidential review]. They include:
- Overview document that describes the project, goals, and weekly essay topics
- Weekly reflection essay prompts
- Supplementary materials used in the weekly essay prompts

*Instructor positionality*
The instructor for this course is a Black female, tenured faculty member. During initial attempts to add equity, diversity, and inclusion (EDI) activities to the course, there was significant student pushback. Students didn't believe that these activities were relevant to the course content. In response to this feedback, she added the APS News activity to frame the EDI activities as coming from the broader physics community, which greatly reduced student pushback. The reason for this is likely because students could not assume that the EDI activities were motivated by the faculty member's identity as a minority woman. Adding the Practicing Professionalism framework seemed to further reduce student pushback, possibly because the entire course is now framed in terms of helping students understand all aspects of what it means to do physics, including considerations around EDI.

**About University of Colorado Boulder**
CU Boulder is a large, public, predominantly white, research intensive institution. The Modern Physics for Engineers course is a sophomore-level course that serves as an introduction to quantum mechanics (QM). For some engineering majors, this course fulfills a degree requirement and is typically the last physics class engineering students will take. The version of the course that we teach is the result of several years of course transformation[1-2] with an emphasis on conceptual foundations of QM as well as real-world applications. The class consists of two 75-minute lecture periods per week with several hours of optional, but encouraged, collaborative help sessions and office hours. The College of Engineering at CU Boulder is 25% female-identifying, 75% male-identifying, and 16% of students are first-generation college students. The racial demographics are: 67% white, 10% Asian, 10% Hispanic/Latino, 8% International, 2% African-American, 1% American-Indian/Alaska Native, <1% Native Hawaiian/Pacific Islander, and 2% Unknown[3]. Our Modern Physics course typically enrolls between 75-120 students each semester, and the student population reflects the demographics of the College of Engineering.

The CU Boulder Physics Department offers two sophomore-level Modern Physics courses---the one for predominantly engineering students that we report on in this paper, and a sister course for physics majors. The emphases of these courses depend on who teaches them, but generally they cover the same content with the physics majors course having a slight emphasis on mathematical formalism and the engineers course having an emphasis on conceptual understanding. Either course satisfies degree requirements for both physics and engineering students. We have implemented the EDI materials in the course for engineers because this is the course we had access to, but there is no reason that the same materials could not be included in the physics majors course.

Course materials for the two-part unit on representation in STEM are available at [link removed for confidential review]. They include:
- Required and optional readings
- Homework and exam questions related to the unit
- Handout given to students with graphs for the data interpretation activity
- Lecture slides associated with the in-class activities

*Instructor positionality*
The instructor for this course is a tenured white male professor who has spent decades working on building equitable and inclusive environments for students. Despite having long-standing interest and commitment to EDI work, conducting this work as part of the formal curriculum for the first few iterations was risky and challenging, and it was only with the assistance and support of co-instructors that we were able to manage the construction and enactment of this unit. The unit on representation was co-designed and co-taught by a white woman, who participated in the course as both a graduate teaching assistant and a postdoctoral guest instructor.

# Appendix B

**Carleton: Supplementary Evidence of Student Outcomes**
Two terms of reflection essays for all students in Atomic and Nuclear Physics have been saved anonymously. The corresponding demographic data for each student is also available and can be tied back to individual essays if need be. Carleton's Science Education Resource Center (SERC) performed a textual analysis of the final reflection essay that asks students:
- Do you agree with your original opinions of the nature of physics?
- What did you find most interesting and surprising? What made the most difference to you personally?
- How has your conception of who a physicist is, what a physicist does, and/or what the physics community broadly cares about changed?

Table 1 provides an excerpt of the codebook with exemplar student quotes.

**Table 1.** Excerpt of the SERC codebook for final reflection essay that focuses on the impacts of the reflection essay assignment on student attitudes towards the physics community.

| Code | Description | Example quote |
|---|---|---|
| Who does physics | Describes an expanded view of who does physics. Includes stating that no one type of person is a physicist. | "I was most surprised by how little I know about what a real physicist looks like today. Not only is my image of a typical physicist biased towards white males, but I did not have a good idea of what jobs people do after graduating with a degree in physics. I have a better idea after some of our reflection essays and discussions in class, but I have realized that the range of who physicists are and what they do is much wider than I initially believed." |
| What physicists care about | Physicists care about a broad range – including impact of their work on society, DEI. May explicitly state a shift in perspective on what the field cares about. | "And finally, my perception of how physicists think and what they care about has changed. I've learned how much more there is to physics than just simply developing models and following the scientific method. Ironically, in the last sentence of my first essay I wrote that a good physicist keeps an open mind, referring to the fact that paradigms can change, but after writing all these essays, I've realized just how narrow minded that was, and that truly keeping an open mind involves having an awareness for a lot more than just data." |
| Reflection Intervention | Explicitly describes how reflection/writing intervention has influenced them | "Because of these reflection essays, I have gained a broader picture of what it means to be a physicist. The physics community is not nearly as isolated from the "real world" as I originally thought, and it is not solely concerned with research and advancing the field. I know that representation in the field is far from perfect, but I take heart in the fact that these conversations are happening and that many of us are aware of how the field could be improved." |

| Self included | Describes seeing themselves as a scientist or physicist | "The most surprising fact I learned doing this project was that not all high schools offer physics. This sucks because some people will never get a chance to see if they have a passion for it... I used to think that Physics was a very elitist field where they only wanted the best and brightest to be involved. But now I learn it is almost the opposite, physics wants to get new minds on board. It also showed to me that if you work hard enough and you want it bad enough your brain will adapt to the challenge and make you smarter. Seeing my hard work pay off has the biggest effect on me overall so the essay that had the most difference on my life was the one where I talked about the imposter syndrome and how I felt I wasn't smart enough to do it. This gives me hope for the future and I can't wait for what is next." |
|---|---|---|

**CU: Supplementary Evidence of Student Outcomes**

As part of the in class discussions on representation in STEM included in the CU Modern Physics course, we engage students in a notecard activity. After reviewing and discussing data on participation in STEM, each student writes an answer to the question "why does representation matter?" on a 3 x 5 index card. Students trade cards with someone on the other side of the room (twice) such that each student ends up with an anonymous card. We then facilitate a whole-class discussion asking students to share and respond to ideas on the card they received. The anonymity of this activity helps students feel more comfortable speaking up in front of the whole class and responding directly to their peers' ideas.

Over two semesters of implementing the notecard activity, we have seen students come up with a variety of responses to the question, "Why does representation (in STEM) matter?" Common themes that arose in the discussions are shown in Table 2, along with an exemplar quote for each theme.

**Table 2.** Common responses from students to the question, "Why does representation matter?"

| Why does representation matter? | Example quote |
|---|---|
| Leads to diverse ideas and perspectives | "Diverse groups of people bring in diverse ideas" |
| Diversity leads to better science | "Incredible breakthroughs in science happen when different people with different backgrounds and experiences approach a problem together" |
| Sense of belonging | "Feeling like you belong (i.e., seeing people who look like you)" |

| Opportunities for individuals | "Representation matters because it helps everyone be able to pursue their passions. When young people see people that look like them in a field they are interested in, they are more likely to find role models/mentors/heroes and think that they can be like them." |
|---|---|
| Equity | "In an ideal world, who/where you're born would have no impact on the opportunities you have, and striving for that is really important." |
| Distribution of power/influence | "Without representation, groups can make decisions or assumptions that can directly affect you without consent or concern." |
| Recognition | "leads to more understanding and recognition for minority groups" |
| Dismantle prejudice | "Representation is important to fight prejudice and old mentalities, like saying 'women aren't fit to be engineers'" |
| Access to education/scientific knowledge | "while the science does not change depending on who does it (ex g=9.8 always), the people that that info reaches is highly impacted by who presents it." |